\documentclass[prd,preprint,tightenlines,floatfix,showpacs,preprintnumbers,nofootinbib,eqsecnum]{revtex4}
 \usepackage[dvips,final]{graphicx}
  \usepackage{amssymb}
   \usepackage{amsmath}
    \usepackage{amsfonts}
     \usepackage{epsfig}
      \usepackage{bm}

\begin{document}

\begin{center}{\Large\bfseries Experimental and theoretical premises of new stable hadron existence} \vskip 5mm Yu. N. Bazhutov$^1$, \framebox[1.1\width]{G. M. Vereshkov}\,, V. I. Kuksa$^2$\vskip 5mm {\small {\it
$^1$ Scientific Research Center of Physical Engineering Problems ``Erzion'',\\
Moscow, Russia, erzion@mail.ru}}\vskip 5mm {\small {\it
$^2$ Theoretical Physics Department, Southern Federal University,\\
Rostov-on-Don, Russia, vkuksa47@mail.ru}}
\end{center}
\vskip 5mm \centerline{\bf Abstract}  The hypothesis of the new stable heavy hadrons existence is proposed which follows from Cosmic rays physics indirect data. It is shown that the
hypothesis does not contradict Cosmochemical data, Cosmological test and the restrictions on New Physics effects. This conclusion is based on the most important property of the new
hadrons - repulsion strong interaction with nucleons at large distance asymptote. This effect is substantiated theoretically in the framework of the low-energy hadron interaction
model. Some extensions of Standard Model is considered where new stable and metastable quarks appear.


\section{Introduction}

The experimental registration of the Higgs boson completed the particles content of Standard Model (SM). However, the SM can not be considered as complete theory, since it contains
unresolved internal problems. Moreover, the existence of the Dark Matter (DM), which is required by astrophysics data, remains the essential phenomenological evidence for New Physics
beyond the SM \cite{01}-\cite{03}. Appropriate particle physics candidate as DM carrier should be stable heavy particles weakly interacting with ordinary matter (WIMP). As a rule, such
particles are considered in the framework of supersymmetric, (hyper)technicolor or Grand Unification extensions of SM.

Indirect experimental indications on the existence of new metastable hadrons appear systematically in Cosmic rays physics \cite{1}-\cite{8b}. Taking all the data into account one may
come to the following preliminary conclusions \cite{8}-\cite{8e}:
\begin{itemize}
\item mass of particle $M > 100\, GeV$;
\item lifetime $\tau > 10^{-5}$ s;
\item electric charge $Q\ne 0$;
\item absorption length $\Lambda \sim 10^4\,g/cm^2$;
\item intensity of hypothetic component of cosmic rays is about $10^3$ times less than intensity of primary component with the same energy spectrum;
\item interactions of new particles with matter (Fe, Pb) generate electromagnetic, nuclear (long range) and so-called ``short range'' showers with neutrons and without penetrating particles.
\end{itemize}
Particles possessing the above mentioned properties, provided they really exist, are new (meta)stable hadrons (NSH) with a somewhat higher (in comparison with ordinary hadrons)
penetrating capability. In the framework of the already existing experimental programs on Cosmic rays physics \cite{9} it is possible to conduct a targeted search for new particles,
including new methods \cite{8,10}. Systematical study of experimental and theoretical aspects of this problem are represented in a series of works, where experimental program search of
new hadrons and their theoretical interpretation were considered \cite{8aa}, \cite{10.1}-\cite{10.6}.  The possibilities of collider experiments also rise now due to their reaching
higher energies. Taking into account these perspectives the hypothesis of NSH existence deserves a more detailed discussion. Let us point out that the hypothesis of NSH existence
usually is considered to be stumbling on numerous experimental and theoretical limitations which impossible to overcome without artificial and unreal arguments. For instance, in
Refs.\cite{10.7}-\cite{10.9} the main problems of 4th generations exstention were considered in application to DM and Higgs boson studies at the LHC. However, in reality, there are
just two critical arguments:
\begin{enumerate}
\item Cosmochemical data \cite{11,12} along with Cosmological test \cite{13} exclude existence of new stable hadrons whose interaction with nucleons has an attracting nature;
\item It is very difficult to conduct inclusion of NSH into the theory of elementary particles without contradicting with experimental data on checking the Standard
 Model.
\end{enumerate}
Let us pay attention to the fact that the indication at the nature of the interaction between new hadrons and nucleons takes a prominent place in the first argument. The purpose of
this study is to show that, strictly speaking, it should be formulated somewhat differently: Cosmochemical data together with the Cosmological test are not inconsistent with the
hypothesis of existence of new neutral stable hadrons, whose interaction with nucleons has a repulsive nature. Accepting this hypothesis one may assume that in Cosmic rays experiments
a charged metastable (with a life-time order of neutron one) isotopic partner of the neutral stable hadron displays itself. Responding to the second critical argument we shall
demonstrate a possibility of coordinating this hypothesis with the theory of elementary particles at the concrete examples, where this quark naturally arises and does not contradict to
SM predictions. For instance, the extensions of SM with singlet quark or some technicolor models with vector-like interactions sytisfy to the restrictions on Peskin-Takeuci (oblique)
parameters and the processes which are caused by FCNC. moreover, it was shown in \cite{13a} that in the case of SM extension with 4-th generation the contribution of new heavy quarks
in vector boson coupling may be compensated by the contribution of 50 GeV neutrino. It should be noted, also, that the last experimental rigid restrictions on cross-section of
spin-independent WIMP-nucleon interaction \cite{13b} does not exclude the existence of new hadrons. These restrictions were oriented on WIMP especially, because of undergroud mounting
XENON1T is located at an average depth of 3600 m water equivalent and effectively cuts off the hadron component of cosmic rays.

The paper was organized as follows. In the second section, we discuss astrophysical data and formulate the hypothesis of new hadrons existence. Quark composition of new hadrons and
their cosmological evolution are considered in the third and fourth section respectively. Effective model of baryon-meson interactions is modified for the of new hadrons in the fifth
section and its repulsive asymptotics is demonstrated in the sixth section. In the seventh section some variants of SM extensions are considered where new stable hadrons naturally
appear. The main conclusions are represented in the eighth section.

\section{Discussion of Astrophysical Data}

Relatively high intensity of the hypothetical component of Cosmic rays excludes a suggestion that new hadrons are being born as the result of the primary component interaction with the
atmosphere. In this case, the production cross-section must satisfy to the $\sigma > 20$ mcb condition that, however, is contradictory to the results of quantitative analysis of the
process by quantum chromodynamics methods ($\sigma\sim 10$ nb, when new hadrons contain quarks, squarks or gluino with mass $m > 100$ GeV). Therefore new hadrons must be present in the
primary Cosmic radiation with $I/I_0 \sim 10^{-3}$ intensity, i. e. have to be stable. Data of Cosmic rays physics allow to suggest that absolutely stable are either new charged
hadrons or new neutral hadrons which together with metastable charged hadrons form isotopic multiplets. As it will be seen from the following the first suggestion will have to be
rejected. The second one is quite admissible, because of recharge cross-section of neutral particles in atmosphere does make up an approximate quantity of 20 mb.

Cosmochemical data \cite{11,12} rigidly limit relative (to protons) concentrations of hypothetic absolutely stable charged particles in a matter:
\begin{equation}\label{2.1}
C_-\leqslant 10^{-12},\quad C_+\leqslant 10^{-22}.
\end{equation}
These data practically exclude positively charged particles - so called ``wild protons''. The second limitation (\ref{2.1}) relates also to any new neutral hadrons provided they may
form stable bound states with light nuclei. It should be admitted that such hadrons are practically excluded. Suggestion of existing new stable negatively charged hadrons provided they
may be absorbed by the Helium atoms is inconsistent with Cosmological test: according to \cite{13}, concentration of such hypothetical hadrons in a matter after surviving
``conditioning'' in the process of the Universe evolution:
\begin{equation}\label{2.2}
C_x\sim \frac{1}{S}(Gm^2_x/\hbar c)^{1/2}\approx 10^{-8}
\end{equation}
with $m_x=100$ GeV (mass of hypothetical particles) and $S \sim 10^{-9}$ (dense entropy to baryon). Note, an account of the experimental research for exited quarks by ATLAS
Collaboration \cite{13c} ($m_x>6\, TeV$ for unstable particles) gives the value $C_x\sim 10^{-6}$. It follows from the incompatibility between (\ref{2.1}) and (\ref{2.2}) that new
hadrons provided they really exist must possess a very specific property: at low energies, asymptote of their strong interactions with usual nucleons must have a repulsive nature. In
this case the hypothesis of existence of new neutral and absolutely stable hadrons with repulsion is compatible to all data (Cosmic rays, Cosmochemical limitations, Cosmological test):
\begin{itemize}
\item lack of its bound states with proton eliminates the ''wild hydrogen'' problem;
\item lack of bound states with Helium provides new particles burning-out at Cosmogonics stages of the Universe evolution (see section 4);
\item recharge of new neutral hadrons, contained in Cosmic rays, in atmosphere will lead to the effects, described in \cite{8}.
\end{itemize}

\section{Quark composition of new hadrons and their interaction with nucleons}

New hadrons under consideration may be constructed with ordinary light quarks $u,d$ and new heavy absolutely stable ``upper-type'' quarks U with electric charge $Q = 2/3$ which belong
to fundamental representation of $SU(3)$ color group. It is a separate non-trivial task to include U -quarks into the particles theory at the same time evading contradictions with the
experimental results on checking the Standard Model (SM). Naimly, the unacceptable Peskin-Takeuchi (PT) parameters values are generated by chiral structure of the new fermionic sector,
while vectorlike interactions provide the agreement with the SM precision data. So, we have to consider new quarks with vectorlike (non-chiral) gauge interactions, which arise, for
example, in the chiral-symmetric models or extension of SM with singlet quark.

Stable and long-lived (with a life-time order of neutron one) new hadrons are divided into three families of particles with character masses of m, 2m and 3m where $m > 100$ GeV is mass
of U-quark. Quantum numbers and quark content of these particles are represented in Table 1.
\begin{center}
{Table 1. Hadron characteristics of U -type}
\end{center}
\begin{center}
\begin{tabular}{||l|l|l|l||}
\hline $J^P=0^-$         &$T=\frac{1}{2}$    &$M=(M^0\,M^-)$   &$M^0=\bar{U}u$,\, $M^-=\bar{U}d$\\ \hline
$J=\frac{1}{2}$   &$T=1$              &$B_1=(B_1^{++}\,B_1^+\,B_1^0)$ &$B_1^{++}=Uuu,B_1^+=Uud,B_1^0=Udd$\\
\hline $J=\frac{1}{2}$    &$T=\frac{1}{2}$     &$B_2=(B^{++}_2\,B^+_2)$ &$B^{++}_2=UUu,B^+_2=UUd$\\ \hline $J=\frac{3}{2}$ &$T=0$ &$(B^{++}_3)$ &$B^{++}_3=UUU$\\ \hline
\end{tabular}
\end{center}
Here, the hadrons $M^0,\,B^+_1,\,B^{++}_2,\,B^{++}_3$ are stable, at that the doublet of particles $(M^0,M^-)$ was named erzion in \cite{8aa}. The states in Table 1 were also
considered in \cite{10.7} and \cite{10.8} where $U$-type quark is contained in 4-th generation.

Interactions between new hadrons and nucleons at low energies are described by a model of the mesonic exchange in terms of a corresponding effective Lagrangian (in analogy with
approach in \cite{14}). Let us point out at the first onset that interactions is non-trivial: between new mesons $M$ and ordinary nucleons $N$ there is a lack of one-pionic exchange as
well as exchanges through all other pseudoscalar mesons. This conclusion is based on spin-isotopic properties of a doublet in the first line of the Table 1: Yukawa bonds of $M^+\tau_a
M\pi_a$ type, which might have provided one-pionic exchange in $MN$-channel, are forbidden by demand of effective Lagrangian's invariance with respect to transformations of the
Poincare group. Vector $\omega,\,\rho$ and scalar $s,\,\delta$ mesons exchange (Born diagrams) and two-pionic exchange (loop diagram) are allowed. Other diagrams either reduce to
renormalizations of effective coupling constants of the meson exchange theory or contains more heavy mesons and do not contribute to asymptote of the $MN$ -potential. This potential
satisfies general demand of the isotopic invariance:
\begin{equation}\label{2.3}
U(\vec{r})=U_1(\vec{r})+\bar{\tau}_M \bar{\tau}_N U_2(\vec{r}),
\end{equation}
where $\bar{\tau}_M,\, \bar{\tau}_N$ are Pauly matrixes, determined in isotopic space of the $M$ and $N$ particles. It will be shown in the sixth section, that the asymptote of $MN$
interaction, determined mainly by the vector exchange, has a repulsive nature for all four pairs of the interacting particles. Thus, $M$ particles are being ejected from nuclear matter
by strong interactions. Due to this reason neutral $M^0$ particles are not being absorbed by nuclei; as to charged $M^-$ particles, they could be absorbed by nuclei due to
electromagnetic interaction. However $M^-$ particles are unstable in a free state, and it is evident that the nature of strong $MN$ interactions inside nucleus excludes appearing
defect of mass, necessary for their stabilization. The charge $M^-$ and neutral $M^0$ particles can manifest themselves in Cosmic rays and as carrier of DM. In the work \cite{8aa}
these particles were considered in the frame-work of Erzion Model (see, olso, the fifth and sixth sections of this work), where neutral particles were called erzion and enion.

Interacting properties of the baryon type $B_1$ and $B_2$ particles (the second and third line in Table 1) are similar to nucleonic one and together with nucleons they may compose
atomic nuclei. It will be demonstrated later that this circumstance does not prevent $B_1$ and $B_2$ burn out in the course of Cosmochemical evolution. There are no problems also with
$B_3$ isosinglet interacting with nucleons mainly through an $\eta$ and $\eta^{'}$ - mesons exchange. The constant of such interaction as it follows from the quark model of the mesonic
exchange (see the fifth section), is not a great one, i. e. $B_3N$ -interaction is suppressed in comparison with $NN$ interaction.

There are another type of hypothetical hadrons which possess analogous properties of strong interactions. They are constructed from stable D quark of the ``down'' type with $Q = -1/3$
electric charge. Quantum numbers and quark content of these particles are represented in Table 2.

\begin{center}
{Table 2. Hadron characteristics of $D$ -type}
\end{center}
\begin{center}
\begin{tabular}{||l|l|l|l||}
\hline $J^P=0^-$         &$T=\frac{1}{2}$    &$M_D=(M^+_D\,M^0_D)$   &$M^+_D=\bar{D}u$,\, $M^0_D=\bar{D}d$\\ \hline
$J=\frac{1}{2}$   &$T=1$              &$B_{1D}=(B_{1D}^+\,B_{1D}^0\,B_{1D}^-)$ &$B_{1D}^+=Duu,B_{1D}^0=Dud,B_{1D}^-=Ddd$\\
\hline $J=\frac{1}{2}$    &$T=\frac{1}{2}$     &$B_{2D}=(B^0_{2D}\,B^-_{2D})$ &$B^0_{2D}=DDu,B^-_{2D}=DDd$\\ \hline $J=\frac{3}{2}$ &$T=0$ &$(B^-_{3D})$ &$B^-_{3D}=DDD$\\ \hline
\end{tabular}
\end{center}
In this table, the states $M^+_D,\,B^0_{1D},\,B^0_{2D},\,B^-_{3D}$ are stable. Particles possessing a similar quark composition appear in various high-energy generalizations of SM, in
which $D$-quark is being originated from a singlet of weak interactions group. For example, each quark-lepton generation in $E(6)\times E(6)$-model contains two singlet $D$-type
quarks; just the same quark appears from the Higgs sector at supersymmetric generalization of $SU(5)$ Great unification model. Experimental consequences of existing singlet $D$-quarks
are discussed, for instance, in \cite{15}-\cite{18} and it is always suggested with reference to Cosmological restrictions that new hadrons are unstable due to mixing of singlet
$D$-quarks with standard quarks of the ``down'' type. There are studies in which singlet $U$-quarks are discussed \cite{19}, however, Cosmological restrictions for them are absent
despite a frequently met opinion: results of the Cosmochemical evolutions of the hypothetical $U$ and $D$-types hadrons with their stability suggested, differ very strongly.

\section{Cosmochemical evolution of new hadrons}

{\bf $\bf U$ -type hadrons.} In principle, astral and planetary matter may contain absolutely stable $M^0,\, B^+_1$ and $B^{++}_2$ particles as well as $B^{++}_3$ and $\bar{B}^{++}_3$.
The antiparticles $\bar{M}^0,\,\bar{B}^+_1$ and $\bar{B}^{++}_2$ are burning out due to interactions with ordinary nucleons:
\begin{equation}\label{4.1}
\bar{M}^0+N\to B^+_1+X,\,\,\,\bar{B}^+_1+N\to M^0+X,\,\,\,\bar{B}_1^{++}+N\to 2M^0+X,
\end{equation}
where $X$ is a totality of leptons and photons in the final state. Reactions (\ref{4.1}) are going on outside as well as inside atomic nuclei. There are no coulomb barriers for them,
therefore $\bar{M}^0,\,\bar{B}^+_1$ and $\bar{B}^{++}_2$ burn out during evolution of the Universe must be total. Other stable particles participate in reactions with annihilation of
new quarks:
\begin{align}\label{4.2}
&M^0+B_3^{++}\to B_2^{++}+X,\,\,\,\,\,M^0+B_2^{++}\to B_1^{+}+X,\notag\\
&M^0+B_1^+\to p+X,\,\,\,\,\,\,\bar{B}_3^{++}+B_3^{++}\to X.
\end{align}
Taking into consideration Cosmochemical restrictions (\ref{2.1}) we should demand that $B^+_1,\,B^{++}_2$ and $B^{++}_3$ particles burn out totally. Here comes a very important
conclusion: baryon asymmetry in new quarks sector has a sign opposite to asymmetry in ordinary quarks sector (quarks $U$ burn out and antiquarks $\bar{U}$ remain). Formation of
residual concentrations of new particles $M^0$ and $B^{++}_3$ goes on in two stages. At first in the course of Cosmological evolution until the moment of ``hardening'', quantities of
concentrations reach $10^{-8}$ in accordance with (\ref{2.2}). After that there are reactions (\ref{4.2}) going on in protoastral matter. We don't have the states like $M^0 p$ (``wild
proton'') and $M^0 He$ (``wild Helium'') here ($M^0$ particle is being rejected from nucleons), therefore (\ref{4.2}) reactions do not have the Coulomb barrier. The Coulomb barrier may
appear for the last reaction in (\ref{4.2}) provided a particle is captured by a nucleus heavier than $He$. However, simple evaluations which are based on the quark model reveal, that
more advantageous reaction in the sense of energy should be reaction
\begin{equation}\label{4.3}
B_3^{++}+N\to 3\bar{M}^0+X,
\end{equation}
which may be going on both outside and inside atomic nuclei. Thus there are no limits for total $U$ -quarks burning out and along with them for all positively charged new hadrons to
the level compatible with (\ref{2.1}). The rest of antiquarks $\bar{U}$ according to (\ref{4.3}) exist inside neutral $M^0$ -particles only. Concentration of these particles in matter
is being determined by the baryon asymmetry in the new quarks sector. There are no rigid experimental restrictions for this quantity.

{\bf $\bf D$ -type hadrons.} In this case $M^+_D,\,B^0_{1D},\,B^0_{2D}$ and $B^-_{3D}$ particles and $\bar{B}^-_{3D}$ antiparticles may be encountered in matter. Analogous to
(\ref{4.1}) reactions look like this
\begin{equation}\label{4.4}
\bar{M}^+_D+N\to B^0_{1D}+X,\,\,\,\bar{B}^0_{1D}+N\to M^+_D+X,\,\,\,\bar{B}_{2D}^0+N\to 2M^+_D+X.
\end{equation}
and the Coulomb barriers to them are absent. Annihilation of new quarks goes through the following channels:
\begin{align}\label{4.5}
&M^+_D+B_{3D}^-\to B_{2D}^0+X,\,\,\,\,\,M^+_D+B_{2D}^0\to B_{1D}^0+X,\notag\\
&M^+_D+B_{1D}^0\to p+X,\,\,\,\,\,\,\bar{B}_{3D}^-+B_{3D}^-\to X.
\end{align}
Taking into account Cosmochemical limitations (\ref{2.1}) it is necessary to demand total burning out of $M^+_D$ and $B^-_{3D}$ particles in reactions (\ref{4.5}). It means that the
baryon asymmetry in new quarks sector should have the same sign that is present in ordinary quarks sector. The situation becomes more complicated by the fact that $B^0_{1D},\,B^0_{2D}$
and $B^-_{3D}$ stable baryons may be absorbed by atomic nuclei, therefore at the Cosmogonical stage of the evolution the Coulomb barrier will appear for all reactions (\ref{4.5}). As
the result the concentration of positively charged particles in matter will be an order of (\ref{2.2}) and that is contradictory to the restriction (\ref{2.1}). Therefore all the
models, containing new $D$ -quarks must have a possibility of their destabilization; for example, at the expense of singlet-doublet mixing with down quarks. So, Cosmochemical and
Cosmological arguments reject new stable hadrons with D-quark only. However, existence of new stable hadrons containing the $U$ -type quark is not forbidden.

\section{Interactions between new hadrons and nucleons}

Interactions between new hadrons and nucleons at low energies can be described by a model of mesonic exchange in terms of a corresponding effective Lagrangian. This Lagrangian is
constructed in close analogy with the gauge model of baryon-meson interactions \cite{14}, It was shown in \cite{14}, that low-energy baryion-meson interactions are described by
$U(1)\times SU(3)$ gauge theory with spontaneous braking of symmetries. Here, $U(1)$ is the group of semy-strong interaction and $SU(3)$ is the group of hadrons unitary symmetry (for
the case of three flavors). The model contains baryon octet $B=\{N,\Lambda,\Sigma,\Xi\}$, gauge field, wich idetified with the nonet of vector mesons $V=\{\rho ,\omega , \phi ,
K^{*}_1\}$, and the nonets of pseudoscalar and scalar mesons. The last is identified with the scalar fields of Higgs type. This dynamical realization of the unitary symmetry is in good
agreement with the low-energy hadron experiments. The model predicts three new mass relations betwwen the masses of vector, scalar and pseudoscalar mesons, which are confirmed with
high accuracy. Besides, the gauge sector of the model describes the decay properties of vector mesons, confirming the gauge interpretation of the vector fields.

Further, we construct an additional sector of this model which includes the triplet of hypothetical particles:
\begin{equation}\label{5.1}
M=(M^0,\,M^-,\,M_s),\,\,\,M^0=\bar{U}u,\,\,\,M^-=\bar{U}d,\,\,\,M_s=\bar{U}s.
\end{equation}
Renormalizable Lagrangian can be represented in the form ($\hbar =c=1$):
\begin{equation}\label{5.2}
L_M=L(M,V^0_{\mu},\hat{V}_{\mu})+L(M,\varphi_0,\hat{\varphi})+L(M,\hat{H}),
\end{equation}
where
\begin{equation}\label{5.3}
L(M,V^0_{\mu},\hat{V}_{\mu})=\vert \partial_{\mu}M-\frac{i}{3}g_1 V^0_{\mu}M-ig\hat{V}_{\mu}M\vert ^2-m^2_MM^+M
\end{equation}
describes the interaction of triplet $M$ with the gauge singlet $V^0_{\mu}$ and octet $\hat{V}_{\mu}=\hat{\lambda}_aV^a_{\mu}/2$ of vector fields ($\hat{\lambda}_a$ are Gell-Mann
matrixes);

\begin{align}\label{5.4}
L(M,\varphi_0,\hat{\varphi})&=\lambda_1 \varphi^2_0(M^+M)+\lambda_2\varphi_0(M^+\hat{\varphi}M)\notag\\
                            &+\lambda_3(M^+M)(\hat{\varphi}\hat{\varphi})+\lambda_4(M^+\hat{\varphi}\hat{\varphi}M)
\end{align}
is interaction of the triplet $M$ with the pseudoscalar singlet $\varphi_0$ and octet $\hat{\varphi}$;

\begin{equation}\label{5.5}
L(M,\hat{H})=h_1(M^+M)(\hat{H}^+\hat{H})+h_2(M^+\hat{H}\hat{H}M)
\end{equation}
is interaction of the triplet $M$ with three triplet of scalar higgs fields which are united into $3\times 3$ -matrix $\hat{H}$ \cite{14}. There is the relation between the constants
$g_1,\,g$ and experimentally defined angle of singlet-octet vector meson mixing \cite{14}:
\begin{equation}\label{5.6}
g_1=\frac{\sqrt{3}}{2} g \tan \theta .
\end{equation}
The value $g^2/4\pi\approx 3.16$ follows from the fitting of decay width of vector mesons. The remaining interaction constants
$\lambda_1,\,\lambda_2,\,\lambda_3,\,\lambda_4,\,h_1,\,h_2$ are phenomenological ones. All these constants describe the strong interactions, so we can make some general considerations
concerning their values. These interactions are caused by the constituent quarks exchanges between hadrons and the constants define probability of such processes. The largest
probability has pair-quark exchange with pion quantum numbers. So, the modules of all dimensionless strong constants are restricted by the value $|g_{\pi}|$ for trilinear couples and
$|g^2_{\pi}|$ for fourth order couples. So, for new coupling constant in (\ref{5.4}) and (\ref{5.5}), we assume that $|\lambda_i|\lesssim g^2_{\pi}$ and $|h_i|\lesssim g^2_{\pi}$.

From the Lagrangian of baryon-meson interaction represented in \cite{14} and from the expressions (\ref{5.3}) - (\ref{5.5}), after the symmetry breaking the physical Lagrangian arises
which we use for calculations of $MN$ -interaction potential:
\begin{align}\label{5.7}
L_{int}&=ig_{\pi}\bar{N}\gamma_5\hat{\pi}N +g_{\omega}\omega_{\mu}\bar{N}\gamma^{\mu}N +g_{\rho}\bar{N}\gamma^{\mu}\hat{\rho}_{\mu}N -g_s s\bar{N}N\notag\\
       &-g_{\delta}\bar{N}\hat{\delta}N +\lambda (M^+M)(\hat{\pi}\hat{\pi})+h_{\delta}M^+\hat{\delta}M+ h_s sM^+M\\
       &+ig_{\omega M}\omega^{\mu}(M^+\frac{\partial M}{\partial x^{\mu}}-\frac{\partial M^+}{\partial x^{\mu}}M)+ig_{\rho M}(M^+\hat{\rho}^{\mu}\frac{\partial M}{\partial
       x^{\mu}}-\frac {\partial M^+}{\partial x^{\mu}}\hat{\rho}^{\mu}M)\notag.
\end{align}
In (\ref{5.7}) $N,\,M$ are isotopic doublets (further we consider two-flavors approach); $\omega_{\mu},\,s=f_0$ are singlets; $\hat{\pi},\,\hat{\rho}_{\mu},\,\hat{\delta}=\hat{a}_0$
are two-row matrixes of isotopic triplets. The constant $g^2_{\pi}/4\pi\approx 14$ is known from the date on $\pi N$ -scattering. Four constants are defined in the model in terms of
gauge constant $g$ and mixing angle $\theta$ by the following expressios:
\begin{equation}\label{5.8}
g_{\rho}=g_{\rho M}=\frac{1}{2}g,\,\,\,g_{\omega}=\frac{\sqrt{3}g}{2\cos \theta},\,\,\,g_{\omega M}=\frac{g}{4\sqrt{3}\cos \theta},
\end{equation}
where the values  $g_{\delta}\approx 1.5,\,g_s\approx 4.8$  follows from the mass splitting in baryon octet and the decay properties of scalar mesons \cite{14}. Three new unknown
constants can be expressed through the parameters of the initial Lagrangians:
\begin{align}\label{5.9}
\lambda &=\lambda_3 +\frac{1}{2}\lambda_4,\,\,\,h_{\delta}=\frac{h_2}{g}m_{\rho},\notag\\
h_s&=\sqrt{2/3}\frac{m_{\rho}}{g}[(h_2+2h_1)(\sin\psi +\frac{1}{\sqrt{2}}\cos \psi)-\\
   &-\frac{h_1}{m_{\rho}}(2m^2_{K_1}-m^2_{\rho})^{1/2}(\cos\psi -\frac{1}{\sqrt{2}}\sin \psi)],\notag
\end{align}
where $\psi$ is mixing angle in the sector of scalar mesons, which is expressed through their masses, $\sin\psi \approx -0.8$. As was noted earlier, between the interaction constants
the following relations are assumed:
\begin{align}\label{5.10}
\vert\lambda\vert &\lesssim g^2_{\pi},\quad \tilde{g}^2_{\delta}\lesssim g^2_{\pi},\quad\tilde{g}^2_s\lesssim g^2_{\pi};\notag\\
\tilde{g}^2_{\delta}&=g\vert h_{\delta}\vert /m_{\rho},\quad \tilde{g}^2_s=g\vert h_s\vert/m_{\rho}.
\end{align}
The definitions of the constants $\tilde{g}^2_{\delta},\,\tilde{g}^2_s$ follows from  (\ref{5.9}) and (\ref{5.10}). These constants define the processes of scalar mesons exchanges in
the boson sector of the model. The same exchanges in the fermion sector are defined by the constants $g_s,\,g_{\delta}$, so it is not excluded that $\tilde{g}^2_s\sim
g^2_s,\,\tilde{g}^2_{\delta}\sim g^2_{\delta}$.

\section{Asymptotics of $MN$ -potential at large distance}

The potential and non-relativistic amplitude of scattering in Born approximation for the case of non-polarized particles are connected by the relation:
\begin{equation}\label{6.1}
U(\vec{r})=-\frac{1}{4\pi^2 m}\int f(\vec{q})e^{i\vec{q}\,\vec{r}}\,d^3q,
\end{equation}
where $m$ is reduced mass of scattering particles, $\vec{q}$ is transition 3-momentum in the center-of-mass system.  In the framework of quantum field theory transition amplitude in
momentum representation $F(p,p^{'})$ is defined through the elements of $S$ -matrix according to the relation:
\begin{equation}\label{6.2}
\langle f\vert\hat{S}\vert i\rangle =\delta (p-p^{'})F(p,p^{'}),
\end{equation}
where $p,\,p^{'}$ are the 4-momenta of particles in the initial and final states, $q=p^{'}-p$ are transition 4-momenta. In the case of fermion-boson scattering, it is convenient to
extract from the amplitude multiplicative bispinor term which contains information about the polarization properties of fermions:
\begin{equation}\label{6.3}
F(p,p^{'})=\Phi(p^{'},p)\bar{\psi}^{\pm}(\vec{p}^{'})\hat{O}\psi^{\mp}(\vec{p}).
\end{equation}
Here, upper and low indexes correspond to scattering on fermion and anti-fermion respectively, operator $\hat{O}$ is defined by the structure of interaction Lagrangian and the function
$\Phi(p^{'},p)$ is calculated according to Feynmann rules. In the near-relativistic approach the expression (\ref{6.3}) can be represented in the form:
\begin{equation}\label{6.4}
F(p,p^{'})=\Phi(p^{'},p) A(\vec{p}^{'},\vec{p})\omega^{'}\omega\,,
\end{equation}
where $\omega^{'},\,\omega$ are spinor components of $N^{'}$ and $N$, $A(\vec{p}^{'},\vec{p})$ is the function which is defined by the operator $\hat{O}$. Matching of the expressions
for cross-section in Quantum Mechanics and Quantum FIelds Theory gives the connection between $f(\vec{q})=f(\vec{p}^{'},\vec{p})$ (here, we take into account, that $p^{'}=q+p$) and
functions in (\ref{6.2})-(\ref{6.4}):
\begin{equation}\label{6.5}
f(\vec{p}^{'},\vec{p})=-2\pi i \Phi(p^{'},p) A(\vec{p}^{'},\vec{p})B(\vec{p}).
\end{equation}
The kinematics function is defined according to the expression:
\begin{equation}\label{6.6}
B(\vec{p})=1+\frac{\vec{p}^2(m^2_1 +m^2_2 -m_1 m_2 )}{2m^2_1 m^2_2}.
\end{equation}
For the case of elastic $MN$-scattering $m_1=m_M,\,m_2=m_N$ and according to (\ref{6.1}) and (\ref{6.5}) we have the expression for potential:
\begin{equation}\label{6.7}
U(\vec{r},\vec{p})\approx \frac{i}{2\pi}(1+\frac{\vec{p}^2}{2m^2_N})\int \Phi(p^{'},p)A(\vec{p},\vec{q})e^{i\vec{q}\cdot\vec{r}}\,d^3q
\end{equation}
The contributions into amplitude $\Phi(p^{'},p)$ are defined by four $t$-channel diagrams with one-meson exchange $s,\delta,\omega,\rho$ and one loop-level diagram with two-pion
exchange. In this case, the functions $A(\vec{p},\vec{q})$ have the forms:
\begin{equation}\label{6.8}
A_{\rho}=A_{\omega}\approx 1-\frac{\vec{q}^2}{8m^2_N},\,\,\,A_{2\pi}=A_{\delta}=A_s=\pm (1-\frac{\vec{p}^2}{8m^2_N}+\frac{\vec{q}^2}{8m^2_N}).
\end{equation}
Rerlativistic corrections in (\ref{6.8}) which contain transition momentum $\vec{q}$, are used for more precise evaluation of interaction radius. Further, we calculate the potential
(\ref{6.7}) for the case $\vec{p}=0$. As a result of calculation, the potential functions can be represented in the rorm which is in accordance witn isotopic invariance:
\begin{align}\label{6.9}
U(M^0p,\vec{r})&=U(M^-n,\vec{r})=U_1(\vec{r})+U_2(\vec{r}),\notag\\
U(M^0n,\vec{r})&=U(M^-p,\vec{r})=U_1(\vec{r})-U_2(\vec{r}).
\end{align}
In (\ref{6.9}) $U_1$ and $U_2$ are defined as follows:
\begin{equation}\label{6.10}
U_1=U_{\omega}+U_s+U_{2 \pi},\qquad U_2=U_{\rho}+U_{\delta}.
\end{equation}
The components of potential in (\ref{6.10}) are defined by the expressions:
\begin{align}\label{6.11}
U_{\omega}&=\frac{g^2 K_{\omega}}{16\pi \cos^2\theta}\cdot \frac{1}{r}\exp(-\frac{r}{r_{\omega}}),\quad U_{\rho}=\frac{g^2 K_{\rho}}{16\pi}\cdot
\frac{1}{r}\exp(-\frac{r}{r_{\rho}}),\notag\\
U_s &=\frac{g_s h_s K_s}{8\pi m_M}\cdot\frac{1}{r}\exp(-\frac{r}{r_s}),\quad U_{\delta}=\frac{g_{\delta}h_{\delta}K_{\delta}}{8\pi
m_M}\cdot\frac{1}{r}\exp(-\frac{r}{r_{\delta}}),\notag\\
U_{2\pi}&=\frac{3m_N}{8\pi m_M}\cdot\frac{\lambda g^2_{\pi}K_{2\pi}}{16\pi^2}\cdot\frac{1}{r}\exp(-\frac{r}{r_{2\pi}}).
\end{align}
The coefficients $K_i$ and $r_i$ in (\ref{6.11}) have the following numerical values:
\begin{align}\label{6.12}
K_{\omega}&=K_{\rho}=0.92,\quad K_s=K_{\delta}=1.6,\quad K_{2\pi}=0.62;\notag\\
r_{\omega}&=\frac{1.04}{m_{\omega}},\,r_{\rho}=\frac{1.04}{m_{\rho}},\,r_s=\frac{0.93}{m_s},\,r_{\delta}=\frac{0.93}{m_{\delta}},\,r_{2\pi}=\frac{1.39}{m_N}.
\end{align}
Thus, the calculations reveal that $U_s,\,U_{\delta},\,U_{2\pi}$ functions contain multipliers of the $m_N/m_M$ - kind, sharply reducing intensities of the corresponding interaction.
Mathematically these multipliers appearance is connected with field operators normalization of pseudoscalar $M$ -particles. These multipliers cannot be eliminated by modifying the
Lagrangian interactions, because relativistically invariant bonds of the scalar and pseudoscalar fields cannot contain uneven number of derivatives on 4-coordinates.

According to the expressions (\ref{6.9})-(\ref{6.11}) the interacting nature (attraction $U < 0$ or repulsion $U > 0$) is determined primarily by signs of coupling constants product
and secondly, by relative quantity of $U_1(\vec{r})$ and $U_2(\vec{r})$ functions. Multiplicative suppression of $s,\,\delta$ and $2\pi$ - exchanges lets disregard their contribution
into asymptote of $MN$ potential with a great mass of $M$ -particles, $m_N/m_M \leq 10^{-2}$. This asymptote is formed by vector exchange for which there is no similar suppressing
multiplier, because of pseudoscalar and vector fields bonds contain a derivative. Moreover, it was demonstrated in \cite{14} that all interactions with vector mesons possess a gauge
nature and corresponding coupling constants are expressed through universal $``g''$-constant and angle $\theta$ of singlet-oktet mixing of vector mesons. As a result of calculations we
get approximate equality:
\begin{equation}\label{6.13}
U_{\omega}(\vec{r})\approx U_{\rho}(\vec{r}),\,\,\,g_{\omega}g_{\omega M}=\frac{g^2}{\cos^2\theta},\,\,\,g_{\rho}g_{\rho M}=g^2.
\end{equation}
As it follows from (\ref{6.9}) and (\ref{6.13}) the asymptote of $MN$ interaction, determined by the vector exchange, has a repulsive nature for all four pairs of the interacting
particles ($U(\vec{r})>0$). Thus, $M$ particles are being ejected from nuclear matter by strong interactions. Due to this reason neutral $M^0$ particles are not being absorbed by
nuclei; as to charged $M^-$ particles, they could in principle be absorbed by nuclei due to electromagnetic interaction. However, $M^-$ particles are unstable in a free state, and it
is evident that the nature of strong $MN$ interactions inside nucleus excludes appearing defect of mass, necessary for their stabilization.

\section{Standard Model Extensions with new heavy quark}

In this section, we note some theoretical suppositions of new hadrons existing and point out corresponding extensions of Standard Model (SM). The relevant extension has to contain
extra heavy quarks with QCD type strong interactions,  which provide the possibility to form new heavy hadrons. The most developed approaches are the theory with singlet quark (for
instance $E(6)$ variant of GUT), the extension with fouth generation and mirror models. Each quark-lepton generation tn the $E(6)$ -theory contains singlet D-type quark and the same
quark appears from the Higgs sector of supersymmetric generalization of $SU(5)$. However, such quarks are unstable due to mixing with the standard ones of D-type and we can not build
new stable hadrons. As was noted earlier, there are studies in which singlet U-type quark \cite{19} is considered, where the phenomenology was analised in detail, but the origin was
not discussed. The extensions of SM with fourth generation and their phenomenology are considered during last decades in spite of strong experimental restrictions, for instance,
following from invizible Z-decay channel, unitary condition for CM-matrix, FCNC etc. The main problem of 4-th generation is the contribution of new heavy quarks to the Higgs decays
\cite{10a}. The contribution of new heavy quarks in vector boson coupling may be compensated by the contribution of 50 GeV neutrino \cite{13a,Ilyin,Novikov}, however, such assumption
looks as artificial. There is a wide class of mirror models, which are discussed during many decades, and futher we consider one of the simplest one.

The model $U(1)\times SU_L(2)\times SU_R(2)\times SU_C(3)$ is the simplest mirror extension of SM with an additional gauge group. It is supposed that the global P-symmetry initially
exists in this model which is broken in low energy range. Such a supposition, however, does not make it possible to choice the set of fundamental fields of the model. In Pati-Salam
model \cite{PatSam} the fermion sector is the simplest one - new quarks and leptons are absent, however, the higgs sector is rather complicated. We consider new version of mirror model
$U(1)\times SU_L(2)\times SU_R(2)\times SU_C(3)$ with extended fermion sector.

The main feature of new version of mirror model (MM) is the reconstruction of P-symmetry by introducing of P-partners in fermion sector. Higgs sector contains standard $SU_L(2)$
doublet $\phi$ and $SU_R(2)$ doublet $\Phi$, which provide an additional degree of freedom for the gauge fields $W_{\mu}$ and $V_{\mu}$. These fields describe standard weak
$SU_L(2)$-interaction of standard fields and supper-weak $SU_R(2)$-interaction of mirror fields. Thus, the standard $\Psi$ and mirror $\bar{\Psi}$ sets of fieds can be represented as
follows:
\begin{align}\label{7.1}
\Psi=&\{f,\psi,W_{\mu},S_{\mu},G_{\mu}\},\,\,\,\mbox{where}\,\,\,f=\{l_L,l_R,q_L,q^+_R,q^-_R\};\notag\\
\bar{\Psi}=&\{F,\Phi,V_{\mu},S_{\mu},G_{\mu}\},\,\,\,\mbox{where}\,\,\,F=\{L_R,L_L,Q_R,Q^+_L,Q^-_L\}.
\end{align}
In (\ref{7.1}) the gauge fields $S_{\mu}$ and $G_{\mu}$ correspond to electromagnetic and strong interactions which are general both for the standard and mirror sectors.

Lagrangian of the MM is represented in the form:
\begin{equation}\label{7.2}
L=L_g+L(\psi,\Phi) + L(f,F) + L(\psi,f;\Phi,F) + L_m,
\end{equation}
where $L_g$ is Lagrangian of the gauge fields which is constructed according to standard rules, $L(\psi,\Phi)$ is Lagrangian of higgs fields:
\begin{align}\label{7.3}
L(\psi,\Phi)&=\vert D_{\mu}\psi\vert^2+\mu^2_L\psi^+\psi-\lambda_L(\psi^+\psi)^2+\vert D_{\mu}\Phi\vert^2\notag\\
            &+\mu^2_R\Phi^+\Phi-\lambda_R(\Phi^+\Phi)^2+h(\psi\psi)(\Phi\Phi).
\end{align}
Here, the last term describes the mixing of usual and mirror scalar fields. Interactions of standard and mirror fermions with the gauge fields are described by the folowing Lagrangian:
\begin{equation}\label{7.4}
L(f,F)=\sum_{f}i\bar{f}\gamma^{\mu}D_{\mu}f +\sum_{F}i\bar{F}\gamma^{\mu}D_{\mu}F.
\end{equation}
Lagrangian $L(\phi, f;\Phi,F)$ describes Yukawa interactions of spinor and scalar fields, and $L_m$ describes phenomenological mass terms which are admissible by the model symmetry.
The structure of covariant differentials in (\ref{7.3}) and (\ref{7.4}) is defined by standard rules, for instance:
\begin{align}\label{7.5}
D_{\mu}q_L &=(\partial_{\mu}+\frac{i}{6}g_1 S_{\mu}-ig_{2L}\hat{W}_{\mu}-ig_3\hat{G}_{\mu})q_L,\notag\\
D_{\mu}L_R &=(\partial_{\mu}-\frac{i}{2}g_1 S_{\mu}-ig_{2R}\hat{V}_{\mu})L_R,
\end{align}
where $g_1,\,g_{2L},\,g_{2R},\,g_3$ are the gauge constants. In the framework of the quantum model, based on the Lagrangian (\ref{7.2}), all constants of coupling are running ones. The
model condition of mirror symmetry provides asymptotic properties of running constants - at large energy the following relation must be fulfilled:
\begin{equation}\label{7.6}
g_{2L}=g_{2R}=g_2,\quad \lambda_L=\lambda_R=\lambda.
\end{equation}
Due to the constants in (\ref{7.6}) do not relate to strong interactions these equalities are fulfilled in tree approximation with good accuracy. Therefore, the violation of $P$ -
symmetry in (\ref{7.2}) is ``soft'', it is caused by the inequality of mass parameters which enter to higgs sector (\ref{7.3}), naimly $\mu^2_R\gg \mu^2_L$. This inequality leads to
the inequality for vacuum condensates $\langle 0\vert \Phi\vert 0\rangle\gg\langle 0\vert \phi\vert 0\rangle$, which leads to $m^2_V\gg m^2_W$.

Lagrangian of Youkawa interactions is as follows:
\begin{align}\label{7.7}
L(\phi,f;\Phi,F)&=-(\bar{l}_L\tilde{\phi}\kappa_L l_R +\bar{q}_L \phi \alpha_L q^+_R +\bar{q}_L \tilde{\phi}\beta_L q^-_R)-\notag\\
                &-(\bar{L}_R\tilde{\phi}\kappa_R L_L +\bar{Q}_R \Phi \alpha_R Q^+_L +\bar{Q}_R\tilde{\Phi}\beta_R Q^-_R)+ \mbox{ý.ñ.}\,.
\end{align}
In (\ref{7.7}) $\kappa_L,\,\kappa_R,\,\alpha_L,\,\alpha_R,\,\beta_L,\,\beta_R$ are numerical matrixes in the generation space,
$\tilde{\phi}=-i\tau_2\phi^+,\,\tilde{\Phi}=-i\tau_2\Phi^+$ are transposed doublets. As in SM, the matrixes $\kappa_L,\,\kappa_R,\,\alpha_L,\,\alpha_R$ can be real, and the matrixes
$\beta_L,\,\beta_R$ are complex Hermitian. From the exppression (\ref{7.7}) after break down of the $SU_L(2)$ and $SU_R(2)$ -symmetries the mass terms of spinor fields are extracted,
and physical states arise as a result of diagonalization. The U-type physical state from doublet $Q$ is just the new heavy quark which forms new hadrons under consideration. The D-type
quark state is the superposition of new and standard quarks fields which is not absolutely stable due to mixing (Corresponding information is contained in the matrixes
$\beta_L,\,\beta_R$).

In the expression (\ref{7.7}) asymptotical $P$ -symmetry means that:
\begin{equation}\label{7.8}
\alpha_L=\alpha_R=\alpha,\quad \beta_L=\beta_R=\beta,\quad \kappa_L=\kappa_R=\kappa .
\end{equation}
In the case (\ref{7.8}), the relations between the masses of mirror particles are the same as for usual ones. However, we should take into account that the orign of Youkawa
interactions (\ref{7.7}) and the corresponding energy scale are unknown. One can olso suggest that the origin of asymmetry $\mu^2_R\gg \mu^2_L$ and interactions (\ref{7.7}) have a
common nature. So, we have no sufficient reasons for validity of relations (\ref{7.8}) and, therefore, we restrict themselves by the suggestion that the masses of mirror quarks and
leptons more less then the masses of super-weak bosons.

The set of fields and symmetries of the model assume the phenomenological mass terms as kinematics couplings between the standard and mirror fields. The last have to be singlets with
respect to $SU_L(2)$ and $SU_R(2)$ groups:
\begin{equation}\label{7.9}
L_m=-(\bar{l}_R mL_L +\bar{q}^+_R m_+ Q^+_L +\bar{q}_R^- m_- Q^-_L )+\mbox{h. c.}.
\end{equation}
In (\ref{7.9}) $m,\,m_+,\,m_-$ are the matrixes in generation space. Kinematics couplings (\ref{7.9}) mix up mutualy mirror worlds and lead to the possibility of mirror particles
decays to the ordinary ones. Note, the mixing should be very weak effect, that is the values $m,m_+,m_-$ are essentially less than the masses of mirror particles. So, all effects
beyond SM predictions, naimely an additional contribution to FCNC and oblique parameters, can be suppressed by the inequality $\mu^2_R\gg \mu^2_L$. This statement has one exception -
the model assumes the possibility of new stable particles with the properties which was discussed in the second section of this work.

Further, we assume that all charge leptons are unstable due to mixing (\ref{7.9}) in the lepton sector, while such a mixing is absent in the quark sector. We, olso, assume that
$m_D>m_U+m_e$ in the first mirror quark generation, that is new quark $U$ is stable and $D$ can be metastable. In this case, the neutral stable and charge metastable hadron states
appear which are represented in the Table 1 (some of these hadrons were called erzion and enion in \cite{8aa}) . As was shown in the second section, the charge asymmetry for new
hadrons should be inverse with respect to one for usual hadrons. This condition can be fulfilled by corresponding choice of free model parameters in mirror sector. In the framework of
erzion model we can describe some peculiarities in the spectrum of cosmic rays (the energy spectrum of vertical muons, hadrons cascades etc. \cite{10.5}).

As the example of the GUT, which includes the model under consideration, we note $SU(8)$-model. Quark and lepton fields in this approach are distributed in one asymmetric and two
fundamental representations of the group $SU(8)$. It is essential that both, usual $u_L,d_L$ and mirror $U^C_R,D^C_R$ quark doublets, equivalently enter to asymmetric representation.
So, the baryon asymmetry arises in accordance with conditions of the phenomenological approach (see the second section).

\section{Conclusion}

The existence of new stable particles in nature will lead not only to altering basic comprehension of the elementary particles theory but also to new prospects in the field of nuclear
technologies \cite{10.1,10.2}. In this study, we tried to demonstrate that the existence of new heavy hadrons does not contradicts to Cosmochemical constraints and Cosmological test.
Moreover, indirect data on the Cosmic ray physics stimulate development of the new stable hadrons hypothesis. Here, we show that new hadrons have to contain heavy quark of U -type
which appears in some extensions of SM. The approach under consideration allows to explain some peculiarities in Cosmic ray physics and sutisfy experimental restrictions on New
Physics.

In this work, we consider the model of effective low-energy interactions of new hadrons with nucleons and demonstrate the repulsive asymptote of this interactions. Such property is
principal in the approach under consideration, because of it allows to satisfy the Cosmochemical constraints and Cosmological test. Some extensions of Standard Model with new heavy
quarks are considered, in particular, the simplest variant of mirror model is described in detailes where proper quark sector appears. Our approach can be applied, olso, to the problem
of hadron Dark Matter and Cold Nuclear Transmutation \cite{10.4}.

\end{document}